\documentclass{aa}  
\usepackage{amsmath, amssymb, graphicx, natbib}
\bibpunct{(}{)}{;}{a}{}{,}
\usepackage{txfonts}

\begin{document}
   \title{Neutral interstellar hydrogen in the inner heliosphere under influence of wavelength-dependent solar radiation pressure}

   \author{S. Tarnopolski
          \and
          M. Bzowski  \inst{}
          }

   \offprints{M.~Bzowski (\email{bzowski@cbk.waw.pl})}

   \institute{Space Research Centre, Polish Academy of Sciences, Bartycka 18A, 00-716 Warsaw, Poland\\
              \email{slatar@cbk.waw.pl,bzowski@cbk.waw.pl}
         }

   \date{}

% \abstract{}{}{}{}{} 
% 5 {} token are mandatory
 
  \abstract
   {With the plethora of detailed results from heliospheric missions such as Ulysses and SOHO and in advent of the first mission dedicated to in situ studies of neutral heliospheric atoms IBEX we have entered the era of precision heliospheric study. Interpretation of these data require precision modeling, with second-order effects quantitatively taken into account.}
   {We study the influence of the non-flat shape of the solar Lyman-$\alpha $ line on the distribution of neutral interstellar hydrogen in the inner heliosphere and assess importance of this effect for interpretation of heliospheric in situ measurements.}
   {Based on available data, construct a model of evolution of the solar Lyman-$\alpha $ line profile with solar activity. Modify an existing test-particle code calculating distribution of neutral interstellar hydrogen in the inner heliosphere to take into account the dependence of radiation pressure on radial velocity.}
   {Discrepancies between the classical and Doppler models appear at $\sim 5$~AU and increase towards the Sun from a few percent to a factor of 1.5 at 1~AU. The classical model overestimates density everywhere except a $\sim 60^{\circ}$ cone around the downwind direction, where a density deficit appears. The magnitude of discrepancies depends appreciably on the phase of solar cycle, but only weakly on the parameters of the gas at the termination shock. For in situ measurements of neutral atoms performed at $\sim 1$~AU, as those planned for IBEX, the Doppler correction will need to be taken into account, because the modifications include both the magnitude and direction of the local flux by a few km/s and degree, which, when unaccounted for, would bring an error of a few degrees and a few km/s in determination of the bulk velocity vector at the termination shock.}
   {The Doppler correction is appreciable for in situ observations of neutral H populations and their derivatives performed a few AU from the Sun.}

\titlerunning{Effects of wavelength-dependent radiation pressure on heliospheric hydrogen }
\authorrunning{Tarnopolski \& Bzowski}

   \maketitle
%
%________________________________________________________________

\section{Introduction}
\label{sec:intro}

After discovery at the end of 1960-ties of a diffuse interplanetary Lyman-$\alpha $ glow \citep{thomas_krassa:71, bertaux_blamont:71}, predicted by \citet{fahr:68} and \citet{blum_fahr:70a} as due to scattering of solar Lyman-$\alpha $ radiation on the neutral interstellar gas flowing through the Solar System, development of models of the distribution of this gas in the heliosphere begun. 

At the very early phase, the influence on neutral heliospheric gas of processes going on at the heliospheric interface was neglected. It was assumed that both the ionization due to solar output and the radiation pressure acting on the H atoms are stationary and spherically symmetric around the Sun and that they fall off proportionally to inverse square of heliocentric distance. It was further assumed that the inflowing neutral gas is monoenergetic, i.e., that before the encounter with the Sun the atoms move with identical, parallel-oriented velocities, identical with the macroscopic bulk velocity of the gas. 

These assumptions formed basis of the first model of density distribution of neutral interstellar gas around the Sun: the purely analytical ``cold'' model \citep{fahr:68, axford:72}. It features axial symmetry about the axis of inflow of the gas and shows either a singularity at the downwind axis, when radiation pressure is too weak to compensate solar gravity, or an empty  ``avoidance zone'' in the downwind region with a paraboloidal boundary surface, when radiation pressure overcompensates solar gravity.

Lifting of the monoenergetic assumption (i.e. allowing for a finite temperature of interstellar gas, high enough to yield thermal velocity comparable to the bulk velocity of the gas), adoption of the distribution function of the gas far away from the Sun (``in infinity'') in the form of a Maxwellian shifted in the velocity space by the bulk velocity vector, and assumption that the gas is collisionless on the distance scale comparable to the size of the heliosphere allowed to use the Boltzmann equation to describe the problem of interaction of neutral interstellar gas with solar environment. Its solution brought the ``hot model'' of the gas distribution in the inner heliosphere \citep{thomas:78, fahr:78, fahr:79, wu_judge:79a, lallement_etal:85b}. This model required numerical integration of the distribution function, but the function itself was still analytical. The ``hot model'' in its various implementations became the canonical model of neutral interstellar gas distribution in the inner heliosphere.

Further development of modeling of the interaction and distribution of neutral interstellar hydrogen near the Sun focused on two main topics. On one hand, a lot of effort was put to understand and simulate processes going on at the boundary between the expanding solar wind and the incoming partially ionized interstellar gas, in the region referred to as the heliospheric interface -- and this issue will not be addressed in the present paper: the reader is referred to recent reviews by \citet{baranov:06a, baranov:06b, izmodenov:06a, izmodenov_baranov:06a}. On the other hand, development of the hot model continued, aimed at a more realistic description of distribution of neutral interstellar hydrogen in the inner heliosphere, suitable for quantitative, and not only qualitative interpretation of heliospheric measurements. 

In the first shot, the assumption that the ionization rate is spherically symmteric was eliminated, when \citet{lallement_etal:85a} described the latitudinal modulation of the charge exchange rate with a one-parameter formula $1 - A\,\sin^2 \phi$, implementing it in the CNRS model of the heliospheric gas distribution. This allowed to vary the equator-to-pole contrast of the ionization rate, but required to keep fixed the width and range of the enhanced ionization band.

A different extension of the hot model was proposed by \citet{rucinski_fahr:89, rucinski_fahr:91}, who realized that the rate of ionization by electron impact is not proportional to inverse square of solar distance. Electron ionization is of particular importance for interstellar helium \citep{mcmullin_etal:04a, lallement_etal:04b, witte:04}; for hydrogen it is noticeable inside a few AU from the Sun, where the density of hydrogen gas is already very much reduced by earlier ionization and solar radiation pressure overcompensating solar gravity. Therefore this aspect of physics of neutral interstellar hydrogen was neglected for quite a while since then, but is reintroduced in the most recent versions of the model \citep{bzowski_etal:08a}.

In the next round of development of the density model, implemented in the Warsaw test-particle code, further on referred to as the Warsaw model, gone was the assumption of invariability of radiation pressure and ionization rate \citep{rucinski_bzowski:95b, bzowski_rucinski:95a, bzowski_rucinski:95b, bzowski_etal:97}. Indeed, both the solar EUV flux \citep{floyd_etal:02a} and the solar wind flux \citep{king_papitashvili:05} vary considerably during the solar cycle. The result is a solar cycle variation of the solar radiation pressure, of the EUV ionization rate, and of the rate of charge exchange between neutral H atoms and solar wind protons. This in turn results in appreciable variations of density and bulk velocity of neutral interstellar hydrogen within a dozen of AU from the Sun. 

The Warsaw time dependent model was fully numerical because not only the integration of the distribution function had to be performed numerically, as in the classical ``hot model'', but also the calculation of the integrand, i.e. of the distribution function itself. At this phase of heliospheric research, in lack of sufficiently long time series of measurements of solar wind speed and density, and of the solar EUV output, substitute idealized models of evolution of these parameters had to be used. 

These aspects of development of heliospheric gas models were discussed in greater detail in a review by \citet{rucinski_bzowski:96}.

In the next move, effects of interaction of the solar wind and interstellar gas in the heliospheric interface were taken into account. Owing to the charge exchange between the atoms of interstellar gas and the heated and compressed plasma in front of the heliopause, the original population of neutral atoms is somewhat cooled and accelerated, and a new population of atoms appears. These new neutral atoms originally inherit the properties of the parent plasma population, but further decouple from this plasma and flow through the heliopause and inside the termination shock of the solar winds. Both populations interact further with the ionized components, exchanging charge with the protons from local plasma. Hence, the processes in the heliospheric interface create a few distinct, collisionless populations of neutral atoms (\citet{osterbart_fahr:92, baranov_etal:91, baranov_malama:93}, as discussed in detail by \citet{izmodenov:00} and \citet{malama_etal:06}). 

From the view point of modeling of the distribution of neutral interstellar hydrogen in the inner heliosphere, the most important aspect of the processes going on in the heliospheric interface is the modification of distribution function at TS. Instead of the shifted Maxwellian with parameters homogeneous in space, \citet{scherer_etal:99} adopted a sum of two Maxwellians, with non-isotropic temperatures, shifted by appropriate bulk velocity vectors. The two components of the new functions had parameters being functions of the offset angle $\theta$ from the upwind direction and corresponded to the two thermal populations (primary and secondary), as predicted by the Moscow Monte Carlo simulation of heliospheric interface. This new version of the Warsaw model was time dependent, but still axially symmetric. At this time, enough measurements had been published to attempt to introduce observations-based models of radiation pressure and ionization rate to the simulations.

Axial symmetry was removed in the next round of the Warsaw model development, when anisotropy of the ionization rate was introduced. \citet{bzowski_etal:01a, bzowski_etal:02} allowed the ionization rate to change as a continuous function of heliographic latitude, with the latitudinal profile of the ionization rate continuously changing with the phase of solar cycle. Hence the ionization field in the model became 2D (keeping the axial symmetry about the solar rotation axis) and time-dependent, and since the gas inflow axis and solar rotation axis are inclined at an angle to each other, the model of neutral hydrogen distribution in the inner heliosphere became 3D and time dependent. 

The most recent extension of the Warsaw model is presented in this paper. We improve on the modeling of radiation pressure acting on individual H atoms, which now is not only a function of time, but also of radial velocity of the atom. In the previous versions of the model it was assumed that the profile of the solar Lyman-$\alpha$ line, responsible for the radiation pressure, is flat. In reality it not only is non-flat, but shows considerable variations depending on the phase of solar activity \citep{lemaire_etal:02, lemaire_etal:05}. 

We take this into account and in Section 2 we develop an observation-based model of evolution of the line profile as a function of the line- and disk-integrated flux. We use this model in a newly-developed code  which simulates density and higher moments of distribution function of neutral interstellar hydrogen in the inner heliosphere. We discuss these calculations in Section 3. With the use of the newly-developed code we assess modifications of the local gas density and of its local flux with respect to the results of models neglecting the dependence of radiation pressure on radial velocity. We also show possible implications for interpretation of heliospheric in situ measurements of neutral atoms and pickup ions. This discussion is provided in Section 4. Section 5  offers a summary of the results.

%__________________________________________________________________

\section{The Doppler model of neutral hydrogen distribution in the inner heliosphere}
\label{sec:model}
   \begin{figure}
   \centering
   \includegraphics[width=8cm]{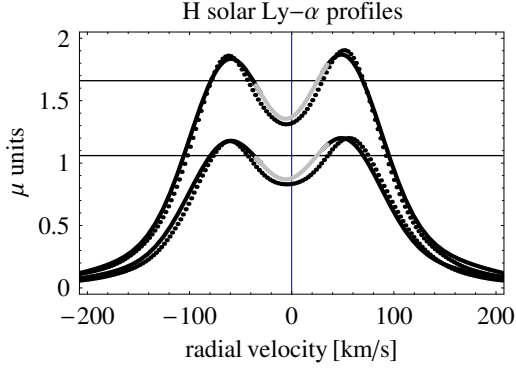}%{aanda07b3_gr1.eps}
      \caption{xa Selected profiles of the solar Lyman-$\alpha $ line observed by SUMER/SOHO (dots) for solar minimum (lower line, Feb. 17, 1997) and solar maximum (upper line, May 20, 2000), compared with the results of the fitted model, specified in Eq. (\ref{eb}) (lines). The gray part of the model lines corresponds to $\pm 30$~km/s around 0 and illustrates the range on the profile which is the most relevant for the thermal populations of neutral H atoms in the heliosphere. The horizontal lines mark the $\mu$ values used in the comparison simulations with the use of the classical hot model.
              }
         \label{xa}
   \end{figure}
   \begin{figure}
   \centering
   \includegraphics[width=8cm]{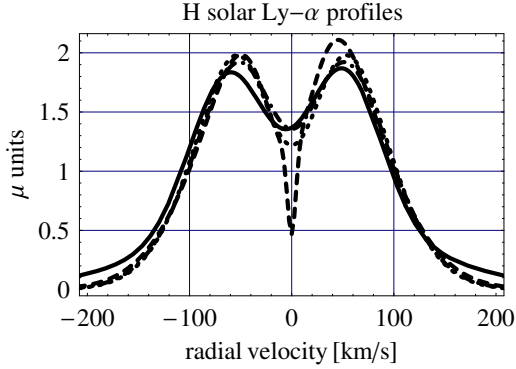}
      \caption{xb Comparison of models of the solar Lyman-$\alpha $ line profile: solid line: the present model, dashed line: Chabrillat \& Kockart, dashed-dot line: Fahr et al.; dots: Scherer.}
         \label{xb}
   \end{figure}

The approach exercised in the present model is a modification of the approach presented by \citet{rucinski_bzowski:95b}. Density and higher moments of distribution function of neutral interstellar hydrogen at a location \vec{R} in the inner heliosphere at a time $\tau$ are calculated by numerical integration of the local distribution function constructed as a product of the distribution function in the source region $w_{\mathrm{src}}$ and of the probability $w_{\mathrm{ion}}$ of survival of the atoms traveling from the source region to the local point  \vec{R} against ionization:
\begin{equation}
f\left(\vec{R}, \vec{V}, \tau \right)=w_{\mathrm{src}}\left(\vec{r}_{\mathrm{src}}\left(\vec{R},\vec{V},\tau \right), \vec{v}_{\mathrm{src}}\left(\vec{R},\vec{V},\tau\right)\right) w_{\mathrm{ion}}\left(\vec{R},\vec{V},\tau \right)
\end{equation}
where $\vec{r}_{\mathrm{src}}(\vec{R},\vec{V},\tau)$ is the start position in the source region of the atom that at the local point \vec{ R} at time $\tau $ has velocity $\vec{ V}$, and $\vec{v}_{\mathrm{src}}(\vec{ R},\vec{ V},\tau $) is relevant start velocity in the source region.

What concerns the code, the distribution function in the source region \(w_{\mathrm{src}}\) a priori can be any reasonable function. In the following we adopted it either as a Maxwellian shifted by the bulk velocity vector, homogeneous in space and independent of time: 
\begin{equation}
w_{\mathrm{src}}\left(\vec{r}_{\mathrm{src}},\vec{v}_{\mathrm{src}}\right)=n_{\mathrm{src}} \left(\frac{m}{2 \pi  k T_{\mathrm{src}}}\right)^{3/2}
\exp \left(-\frac{m\left(\vec{v}_B-\vec{v}_{\mathrm{src}}\right)^2}{2 k T_{\mathrm{src}}}\right)
\end{equation}
or, in the two-populations case, as a sum of two such functions with different parameters $n_{\mathrm{src}}$ (density in the source region), $\vec{v}_B$ (bulk velocity at the source region), and $T_{\mathrm{src}}$ (temperature at the source region); $m$ is the atom mass, and $k$ the Boltzmann constant. Basically, the parameters of these populations are functions of the location in the heliospheric interface \citep{izmodenov_etal:01a}. Since our goal in this paper is investigating the effect of the $v_r$-dependence of radiation pressure on the gas distribution, we left this problem out of the analysis. Some insight into modifications of the gas density profile due to the angular gradient of the parameters of the H populations at the termination shock is provided by \citet{bzowski_etal:08a}.

The challenging issue in the calculation of the local distribution function of the gas is finding of the link between the local velocity of the test atom $\vec{ V}$ at time $\tau $ and location $\vec{R}$ and its position $\vec{r}_{\mathrm{src}}$ and velocity $\vec{v}_{\mathrm{src}}$ in the source region in the scenario where the radiation pressure is a function of time and radial velocity. Indeed, the force acting on a H atom in the heliosphere is composed of the attracting solar gravity and the repelling radiation pressure, which is a function of the spectral flux $I_{\lambda}$ relevant for a given radial velocity $v_r$. The spectral flux $I_{\lambda}$ is also a function of the line-integrated flux $I_{\mathrm{tot}}\left(t\right)$, which varies on time scales from days to dozens of years. Thus the spectral flux responsible for instantaneous radiation pressure acting on a H atom moving with radial velocity $v_r$ is a function of $v_r$, $I_{\mathrm{tot}}\left(t\right)$ and, implicitly, of $t$: $I_{\lambda} = I_{\lambda}\left(v_r, I_{\mathrm{tot}}\left(t\right)\right)$. 

 Since both $I_{\mathrm{ tot}}$ and solar gravity fall off with heliocentric distance as $1/r^2$, the radiation pressure can be expressed by the factor $\mu$ of compensation of solar gravity. In the modeling, this factor is a function of radial velocity and of the line-integrated flux, which in turn is a function of time. Hence finally $\mu = \mu (v_r,t)$. 

The link between the local position and velocity vectors $\vec{R}, \vec{V}$ and the corresponding position and velocity vectors at the source region $\vec{r}_{\mathrm{src}}$, $\vec{v}_{\mathrm{src}}$ is determined by numerical solving of the equation of motion specified in Eq.(\ref{ea}), performed backwards in time. Along with the calculation of the trajectory of the atom, its probability of survival is calculated. 
\begin{equation}\frac{\mathrm{d}}{\mathrm{d} t}\left(
\begin{array}{c}
 x \\
 y \\
 z \\
 v_x \\
 v_y \\
 v_z \\
 u \\
 u_{\beta }
\end{array}
\right)=\left(
\begin{array}{c}
 v_x \\
 v_y \\
 v_z \\
 -\frac{G M\left(1-\mu \left(t,v_r\right)\right)}{r^3}x \\
 -\frac{G M\left(1-\mu \left(t,v_r\right)\right)}{r^3}y \\
 -\frac{G M\left(1-\mu \left(t,v_r\right)\right)}{r^3}z \\
 \frac{1}{r^2} \\
 u\, \beta_{0} (r, t,\phi )
\end{array}
\right)
\label{ea}
\end{equation}
In the equation above, $\beta_{0}(r,t,\phi )=r^{2} \,\beta(r,t,\phi )$ is the scaled ionization rate at a heliographic latitude $\phi $, $r^2=x^2+y^2+z^2$, $G M$ is solar mass times constant of gravity, and $v_{r}(t) = (\vec{r}(t)\cdot \vec{v}(t))/|\vec{r}(t)|$ is radial velocity of the atom. The position vector $\vec{ r}(t)$ has coordinates $x, y, z$ and velocity vector $\vec{v}(t)$ coordinates $v_x,  v_y, v_z$; $u_{\beta }$ is used to calculate the survival probability as 
\begin{equation}
w_{\mathrm{ion}}=\exp \left(-u_{\beta }\right).
\end{equation}
Once the position and velocity vectors of the test atom in the source region are found (respectively, $\vec{r}_{\mathrm{src}}$, $\vec{v}_{\mathrm{src}}$), the value of the  distribution function in the source region $w_{\mathrm{src}}$ can be calculated. The value of the local distribution function is then be obtained as a product of $w_{\mathrm{src}}$ and of the the survival probability $w_{\mathrm{ion}}$. To obtain density and higher moments, the local distribution function is numerally integrated in the velocity space $\mathrm{d}^3V$. 

The model of radiation pressure \citep{tarnopolski:07} was developed based on a series of observations of the solar Lyman-$\alpha $ line profiles, performed by SUMER/SOHO between solar minimum and maximum \citep{lemaire_etal:02} and available on the Web. Since the SUMER instrument is located at the SOHO spacecraft orbiting around the L1 Lagrange point between the Earth and the Sun \citep{wilhelm_etal:99}, the observations are free from geocoronal contamination that affected earlier measurements carried out from low-Earth orbits \citep[e.g. OSO-5,][]{vidal-madjar:75}. The absolute calibration of the wavelength reported by \citet{lemaire_etal:02} is better than $\pm$0.0015~nm. The calibration of intensity was performed by direct comparison of results of integration of the observed profiles with the absolute fluxes obtained with the use of the SOLSTICE experiment. The accuracy of the net flux was reported at $\pm 10$\% level. It seems, however, that the data published on the Web are a smoothed version of actual measurements because despite the accuracy of 10\% reported in the paper the profiles do not show any scatter.

Fitting the functional forms of the profiles used in earlier studies \citep{fahr:78, chabrillat_kockarts:97, scherer_etal:00} did not yield sufficient accuracy of the model (cf Fig.~\ref{xa} and \ref{xb}). To develop the model of $\mu(v_r, I_{\mathrm{tot}})$ discussed in this paper, the original data were rescaled to the spectral flux in the $\mu $ units as a function of radial velocity in km/s. A satisfactory result could be obtained with a 9-parameter function in the form \citep{tarnopolski:07}:
\begin{eqnarray}
	\label{eb}
%\mu \left(v_{r}, I_{tot}\right) & = & \left(A + B I_{tot}\right) \exp \left(-C v_{r}^{2}\right) \Bigl[D + E \exp \left(-F v_{r} - G v_{r}^{2}\right) + {}\Bigr. \nonumber \\
%	& & \Bigl. {}+ H \exp \left(-P v_{r} - Q v_{r}^{2}\right)\Bigr]
\mu \left(v_r, I_{\mathrm{tot}}(t)\right)&=&A\left[1 + B\, I_{\mathrm{tot}}(t)\right] \exp \left(-C v_r^2\right)\\ &\times &\left[1 + D \exp \left(F v_r
- G v_r^2\right) + H \exp \left(-P v_r - Q v_r^2\right)\right] \nonumber
\end{eqnarray}
with the following parameters:

 $\begin{array}{lllllllll}
 A  =  2.4543 \times 10^{-9}, & B =  4.5694\times 10^{-4}, & C =  3.8312\times 10^{-5}, \\ D =  0.73879, & F =  4.0396\times 10^{-2}, &  G =  3.5135\times 10^{-4}, \\ 
 H =  0.47817, &  P =  4.6841\times 10^{-2}, &  Q =  3.3373\times 10^{-4}.
\end{array}$

An interesting aspect of the model is that its sole solar-related parameter is the line-integrated flux in the Lyman-$\alpha$ line $I_{\mathrm{tot}}$. Any dependence of radiation pressure on time goes to the model via $I_{\mathrm{tot}}$. Hence, with a model of behavior of $I_{\mathrm{tot}}$ in time on hand (such as, e.g., discussed by \citet{bzowski:01a} and \citet{bzowski_etal:08a}) we can immediately construct a model of radiation pressure that will be dependent on radial velocity of the atoms as well.

The function defined in Eq.(\ref{eb}) with the parameters as listed above reproduce well the observed profiles both for solar minimum (the profiles from July 27, 1996 to August 24, 1997) and for solar maximum (the profiles from August 20, 1999 to August 22, 2001). The accuracy of the fit for the most interesting region $\pm$140 km/s about the line center exceeds the accuracy of observations declared by the authors. Furthermore, the line-integrated fluxes obtained from the model profiles agree well with the line-integrated data and with the total fluxes reported by \citet{tobiska_etal:00c} for a given day. The differences are on the order of 5\%; only for the profile observed on May 20, 2000 the difference is 9\%, which is still within the absolute calibration accuracy of 10\%. A comparison of the experimental data with the fitted model is shown in Fig. \ref{xa}. A comparison of this model with earlier models \citep{fahr_etal:81a, chabrillat_kockarts:97,scherer_etal:00} is presented in Fig. \ref{xb}.

As the absolute calibration of the line-integrated flux of the Sun changed since the time of publication, we show the original profiles recalibrated by multiplication by appropriate factor so that their absolute flux is in agreement with the values accepted nowadays. The differences between the models in the velocity range $\pm$100 km/s around the line center are on the order of 10 to 15\%. For higher velocities the percentage differences between the models are not so important because the absolute values of radiation pressure are anyway small, and furthermore the fast atoms are not very sensitive to radiation pressure in general \citep{bzowski:08a}. The highest differences in the absolute terms occur at the peaks of the profile, i.e., at the velocities characteristic for the fast wing of the distribution function of thermal populations of neutral hydrogen in the inner heliosphere. With as many as 9 data sets available, covering the whole span of solar activity level, we were able to come up with a model that seems to better reproduce physical reality than the former ones. 

This scheme was devised to enable calculating the density and higher moments of the distribution function in the case of a fully 3D and time dependent radiation pressure and ionization rate. The code has already been used in its full model by \citet{tarnopolski_bzowski:08a} to assess the flux of interstellar deuterium at the Earth orbit and by \citet{bzowski_etal:08a} to investigate the density of neutral interstellar H at the termination shock based on the observations of hydrogen pickup ions by Ulysses. In the present study the code was restricted to the spherically symmetric and stationary case to facilitate assessment of the influence of the $v_r$-dependence of radiation pressure on the distribution of interstellar hydrogen in the inner heliosphere, unblended with other departures from the classical hot model.

\section{Calculations}
\label{sec:calcu}
\begin{figure*}
   \centering
   \includegraphics[width=8cm]{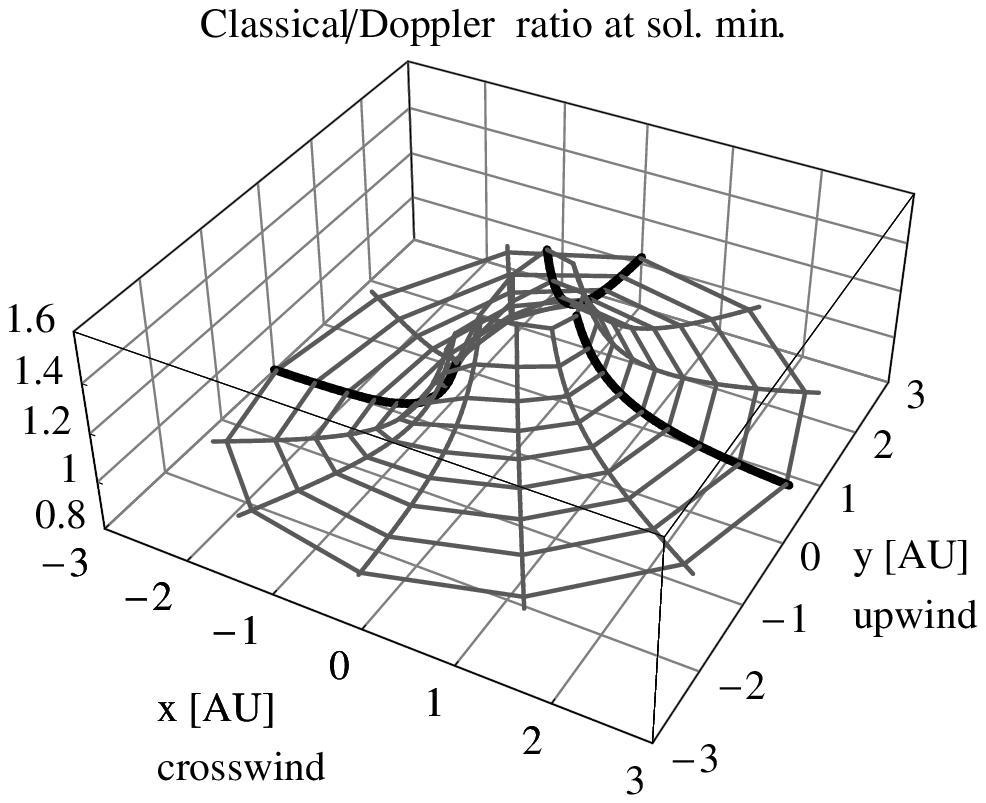}
   \includegraphics[width=8cm]{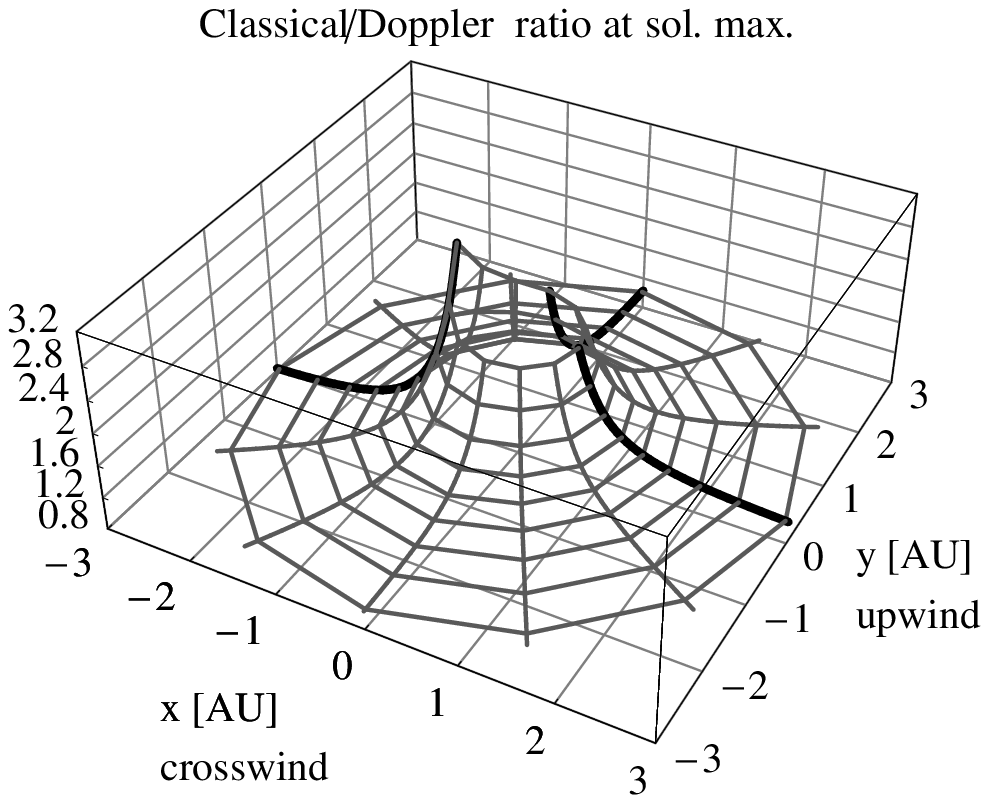}
   \caption{xg Classical/Doppler density ratios in the composite model for the solar minimum (left panel) and solar maximum conditions (right panel). The Sun is at the center in the (0,0) point, the upwind -- downwind direction is along the $y=0$ line. The upwind, crosswind, and downwind axes are marked with heavy lines.}
              \label{xg}
    \end{figure*}
   \begin{figure*}
   \centering
   \includegraphics[width=8cm]{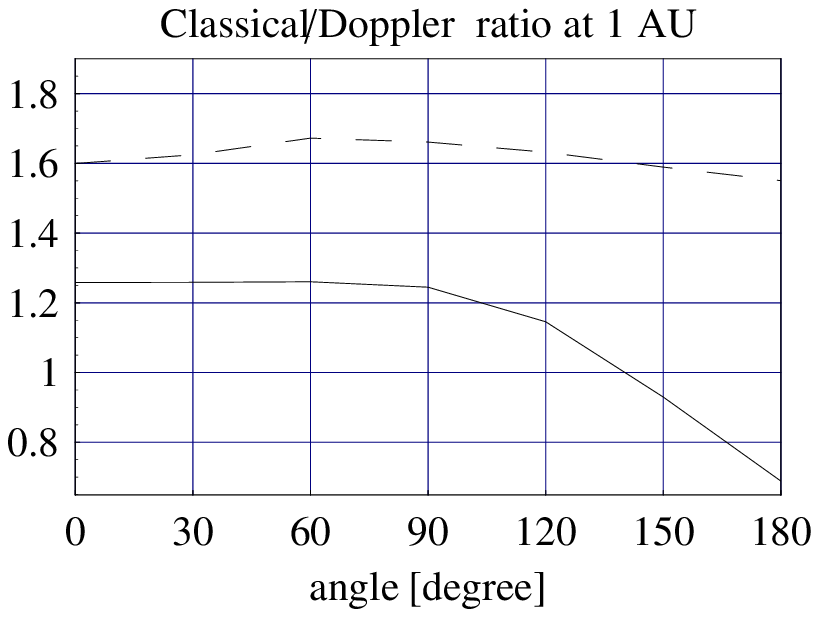}
   \includegraphics[width=8cm]{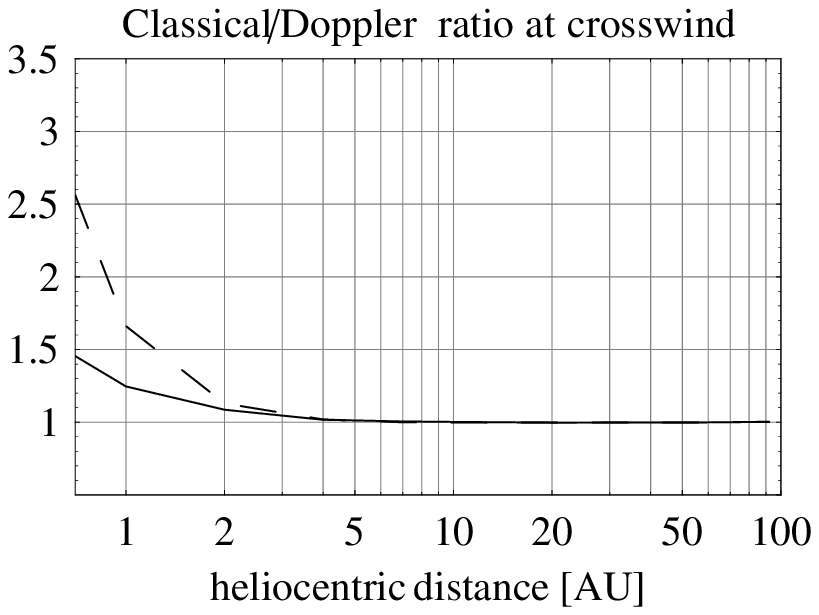}
   \caption{xc Density excess of neutral interstellar hydrogen in the inner heliosphere, defined as a ratio of results of the classical hot model to the results of Doppler model. Left panel: density ratio at 1~AU as a function of offset angle from the upwind direction, right panel: density ratio at crosswind as a function of heliocentric distance. Solid lines correspond to solar minimum conditions, broken lines -- to solar maximum conditions. }
              \label{xc}%
    \end{figure*}

The simulations were performed in two groups. In the first one, we investigate the differences between the classical hot model and the Doppler model in a scenario with two populations at the termination shock: the primary interstellar population and the secondary population, which comes up up between the heliopause and the bow shock from charge exchange between atoms from the original interstellar populations and the local heated and compressed plasma. These results are presented in Section \ref{subsec:twopopul} and will be referred to as the composite model, in the sense that it is composed of the two populations: primary and secondary. In the second group of simulations, we check the robustness of the conclusions drawn in Section \ref{subsec:twopopul} against uncertainties of the bulk velocity and temperature of interstellar gas at the termination shock and against uncertainties in the ionization rate in the inner heliosphere. The results of this series are discussed in Section \ref{subsec:paramscan}.

The calculations were performed on a mesh of heliocentric distances $r$ equal to 0.4, 0.7, 1, 2, 4, 7, 10, 20, 50, and 100 AU and of offset angles from the upwind directions $\theta $ equal to $0\degr$, $30\degr$, $60\degr$, $90\degr$, $120\degr$, $150\degr$, and $180\degr$, separately for solar minimum and solar maximum conditions, characterized by the following values of the line-integrated solar Lyman-$\alpha $ flux and net ionization rate:\\
solar max.: $I_{\mathrm{tot}}=5.50\times 10^{11}$~photons~cm$^{-2}$ ~s$^{-1}$, $\beta = 8 \times 10^{-7}$~s$^{-1}$\\
solar min.:  $I_{\mathrm{tot}}=3.53\times 10^{11}$~photons~cm$^{-2}$ ~s$^{-1}$, $\beta = 5 \times 10^{-7}$~s$^{-1}$.

The boundary conditions of the composite simulation were the parameters returned by the Moscow MC model at the nose of the termination shock for the following set of LIC conditions  \citep{izmodenov_etal:03a}:\\ 
proton density $n_{p,\mathrm{LIC}}=0.06$~cm$^{-3}$, neutral H density $n_{H,\mathrm{LIC}}=0.18$~cm$^{-3}$, temperature of the gas $T_{\mathrm{LIC}}=6400$~K, bulk velocity $v_{B,\mathrm{LIC}}=26.4$~km/s, the density of He$^+$ in the LIC $n_{\mathrm{He}^+} = 0.008$~cm$^{-3}$. 

The resulting parameters of the primary and secondary populations at the nose of the termination shock were the following:

Primary: $n_{\mathrm{TS},\mathrm{pri}}=0.03465$~cm$^{-3}$, $v_{\mathrm{TS},\mathrm{pri}}=28.512$~ km/s, $T_{\mathrm{TS}, \mathrm{pri}} =
6020$~K;\\
Secondary: $n_{\mathrm{TS},\sec }=0.06021$ cm$^{-3}$, $v_{\mathrm{TS},\sec }=18.744$~km/s, $T_{\mathrm{TS}, \sec }= 16300$~K.\\
These parameters were adopted as boundary conditions for the composite simulations with the use of the Warsaw model.

In addition to the dependence of radiation pressure on $v_r$, the code included also a simplified model electron ionization rate, as adopted by \citet{bzowski_etal:08a}. The model is spherically symmetric and identical for solar minimum and solar maximum conditions. Inclusion of this model is a departure from the approach adopted in this study to keep all aspects of the modeling as close to the classical hot model as possible. We decided to include the electron ionization to make sure that we do not miss possible disturbances of the Doppler-to-hot model ratios due to the extra ionization term which does not conform with the $1/r^2$ distance profile of the charge exchange and photoionization rates. 

The simulations were repeated with the use of a model that differs from the former one only by the lack of sensitivity of radiation pressure to $v_r$, i.e. effectively using the classical hot model with the electron ionization term added. The radiation pressure was equal to the spectral flux averaged by $\pm 30$~km/s about 0, as indicated in Fig.\ref{xa}. The numerical accuracy of the solutions was at the level of 1 to 2\%. 

The simulations discussed in Section \ref{subsec:paramscan} were performed assuming there is only one population of interstellar hydrogen at the termination shock. We explored the parameter space adopting the net ionization rates at 1~AU $\beta$, bulk velocities $v_{\mathrm{src}}$, and temperatures $T_{\mathrm{src}}$ covering the entire range of expected values from the low values of $\beta$ relevant for the fast solar wind conditions and solar minimum to the high values relevant for slow solar wind conditions and solar maximum, the ranges of $v_{\mathrm{src}}$ and $T_{\mathrm{src}}$ from the extremes predicted for the primary and secondary populations at the termination shock. 

This scan over $\beta$ was performed assuming the gas temperature and velocity at the termination shock are equal, respectively, to 12500~K and 22~km/s, as derived by \citet{costa_etal:99} from an analysis of heliospheric Lyman-$\alpha$ glow in the upwind hemisphere, for $I_{\mathrm{tot}}$ relevant for solar minimum and maximum conditions(see above). The scan over $v_{\mathrm{src}}$ and $T_{\mathrm{src}}$ space was performed assuming $\beta$ and $I_{\mathrm{tot}}$ values as listed for solar minimum and maximum conditions.

 The baseline set of simulations was performed with the use of the newly-developed Doppler model. Comparison simulations were performed with the use of the classical hot model on identical spatial mesh and assuming identical parameters at the termination shock and identical ionization rate. In this section the electron ionization was not included and hence the ionization rate was falling off as $1/r^2$. 

\section{Results}
\label{sec:results}

\subsection{Composite model}
\label{subsec:twopopul}

\begin{figure}
   \centering
   \includegraphics[width=8cm]{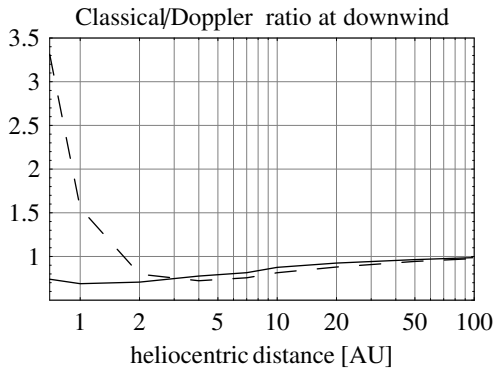}
   \caption{xd Density ratio of the classical hot model to Doppler model at the downwind axis as a function of heliocentric distance. Solid line: solar minimum conditions, broken line -- solar maximum conditions. }
              \label{xd}
    \end{figure}

The simulations show that neglect of the $v_r$-dependence of radiation pressure produces an excess of local hydrogen density (see Fig. \ref{xg}). The excess $q$ is defined here as the ratio of density returned by the classical hot model $n_{\mathrm{class}}$  to the density returned by the Doppler model $n_{\mathrm{dopp}}$: $q = n_{\mathrm{class}}/n_{\mathrm{dopp}}$. Discrepancies between the classical and Doppler models appear at $\sim 3$~AU and increase towards the Sun. Depending on the phase of the solar cycle and the offset angle from upwind, at 1~AU they are from a few dozen percent up to a factor of $\sim 1.7$ (see Fig. \ref{xc}). The excess density is a strong function of solar activity and while during solar minimum its typical values at 1~AU are about 25\%, they are a little more than twice as large at solar maximum. Exception is a cone region around the downwind axis with the opening angle of $\sim 60\degr$, as shown in Fig. \ref{xc} and \ref{xd}: at the downwind axis the hot model predicts a density deficit, which almost does not depend on the activity phase except at closest distances to the Sun. During solar maximum the excess is predicted even in the downwind region, though only for distances smaller than $\sim 1.5$~AU.

\begin{figure*}
   \centering
   \includegraphics[width=8cm]{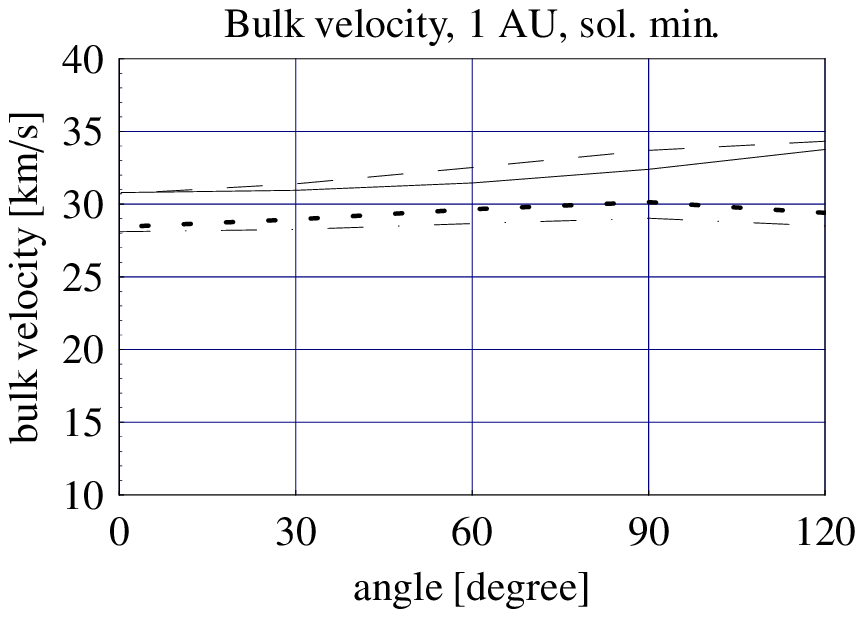}
   \includegraphics[width=8cm]{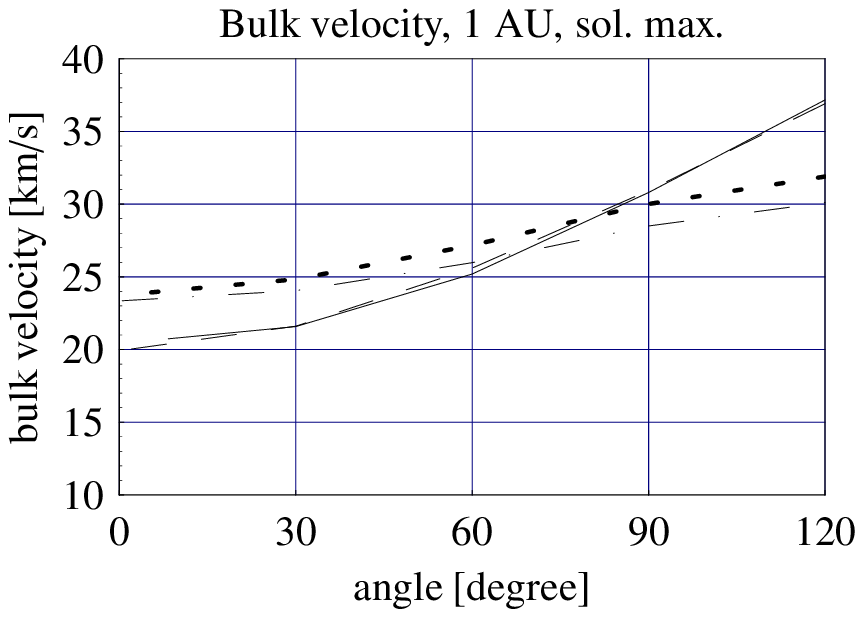}
   \caption{xl Bulk velocities of the primary and secondary populations of neutral interstellar H at 1~AU, calculated using the Doppler and classical models, shown as a function of offset angle from upwind for the solar minimum (left panel) and solar maximum conditions (right panel). Primary population: solid line (classical) and broken line (Doppler); secondary population: dotted line (Doppler) and dash-dot (classical). }
              \label{xl}
    \end{figure*}
The existence of the density excess can easily be understood after inspection of the shape of the solar Lyman-$\alpha$ line profile, shown in Fig. \ref{xa}. Since the profile has a minimum close to  0 radial velocity, the H atoms before approaching the Sun sense a stronger repulsion than in the case when one adopts a ``flat'' radiation pressure with the mean value averaged between $\pm 30$ km/s around the line center. Hence in the Doppler case a larger portion of these atoms will be slowed down and repelled from the Sun, which reduces the density in comparison with the classical case. The ensemble of surviving H atoms has a lower bulk velocity than predicted by the classical model, which leads to a further enhancement of the excess, because the ionization has more time to eliminate the atoms traveling with a smaller speed. On the other hand, there exists a sub-population of atoms which on approach towards the Sun experience a lower radiation pressure in the Doppler model than in the classical model. These are the atoms from the slow wing of the distribution function. In this case, although the Doppler model predicts less deceleration than the classical model, these atoms contribute comparatively little to the entire ensemble because they are more readily eliminated by ionization than the faster atoms. 

The Doppler model predicts different ratios of densities between pairs of offset angles $\theta_1$, $\theta_2$ (e.g. upwind/crosswind, upwind/downwind etc.) than the predictions of the classical hot model. The discrepancies are not strong in the upper hemisphere, but escalate with the increase of difference $\theta _2-\theta _1$. This must be one of the reasons for which interpretation of downwind observations of the heliospheric Lyman-$\alpha$ glow has always been more challenging than of the upwind ones and usually returned different conclusions \citep[e.g.][]{lallement_etal:85b, quemerais_etal:92a}. The challenge can be better appreciated when one realizes that the wavelength-sensitivity effects are convolved with the effects related to variations of solar Lyman-$\alpha$ flux and of the ionization rate, as discussed, e.g., by \citet{rucinski_bzowski:95a, bzowski_rucinski:95a, bzowski_rucinski:95b, bzowski_etal:02}.
\begin{figure*}
   \centering
   \includegraphics[width=8cm]{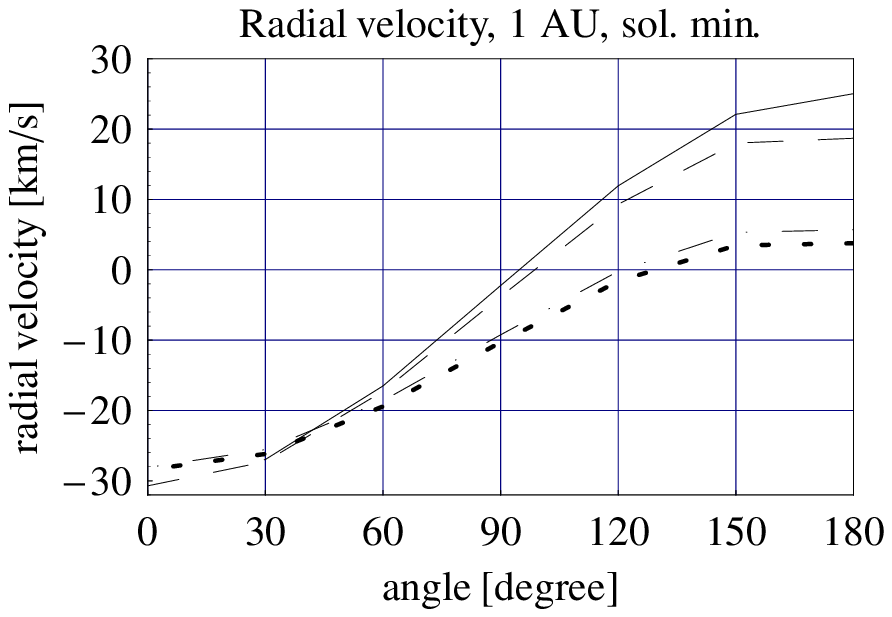}
   \includegraphics[width=8cm]{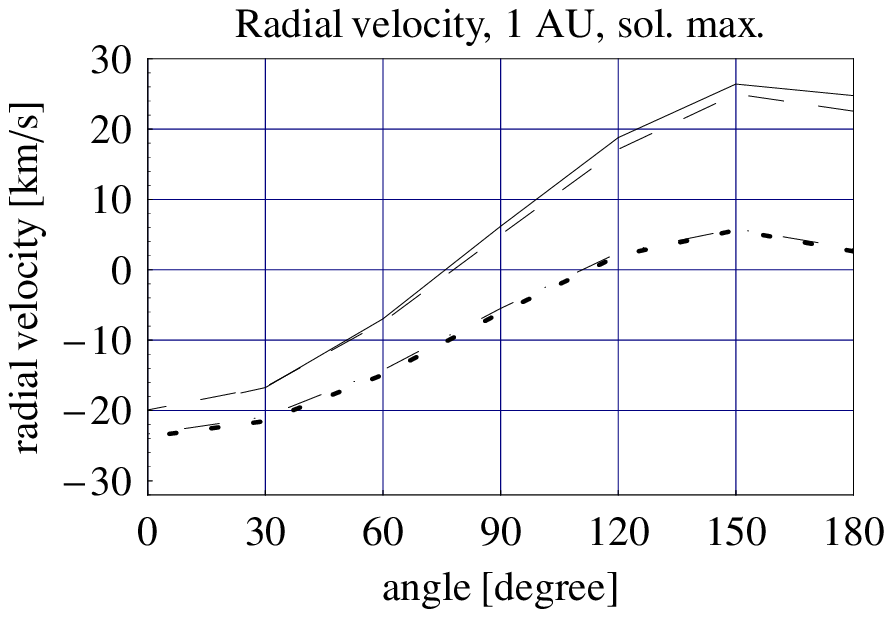}
   \caption{xi Radial velocities of the primary and secondary populations of neutral H atoms at 1~AU as a function of offset angle from upwind for solar minimum (left panel) and maximum conditions (right panel), calculated with the use of Doppler and classical hot model. The upper pairs of lines in both panels correspond to the primary population and the lower pairs to the secondary population. The classical hot model results are drawn, correspondingly, with solid and dash-dot lines, and the Doppler model results with broken and dotted lines.}
              \label{xi}
    \end{figure*}

The Doppler and classical models predict surprisingly small differences between the local bulk velocities of the gas (Fig. \ref{xl}). During solar minimum they are about 1~km/s and during solar maximum they increase to $\sim 2$~km/s. Differences in the radial component of the bulk velocities at 1~AU are also small (Fig. \ref{xi}). They start only at the offset angle $\theta \simeq 60\degr$ and increase towards downwind; the classical model systematically predicts somewhat larger values. Since the differences in $v_r$ are small, we do not expect big differences in model spectral profiles of the heliospheric glow.

\begin{figure*}
   \centering
   \includegraphics[width=8cm]{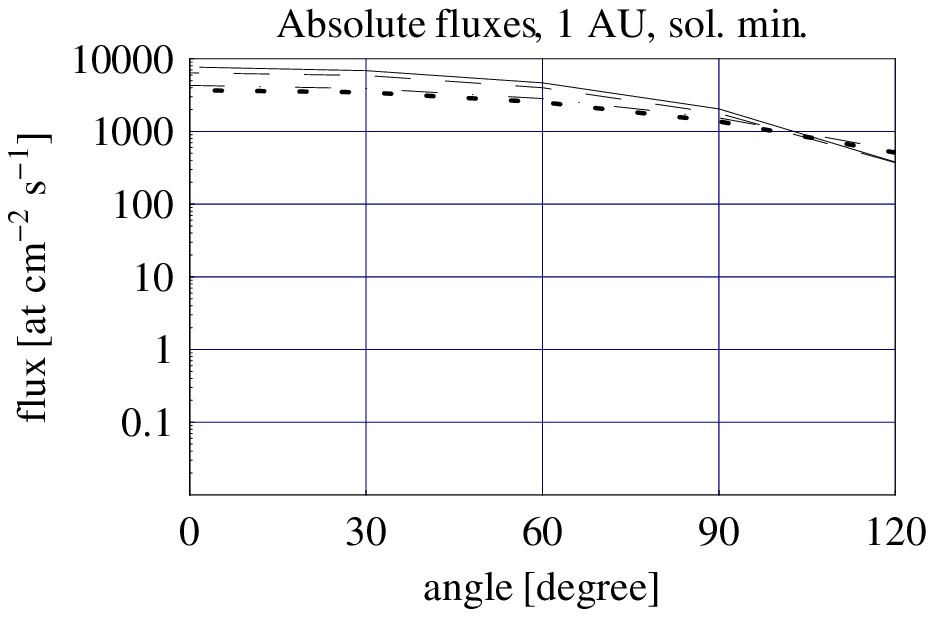}
   \includegraphics[width=8cm]{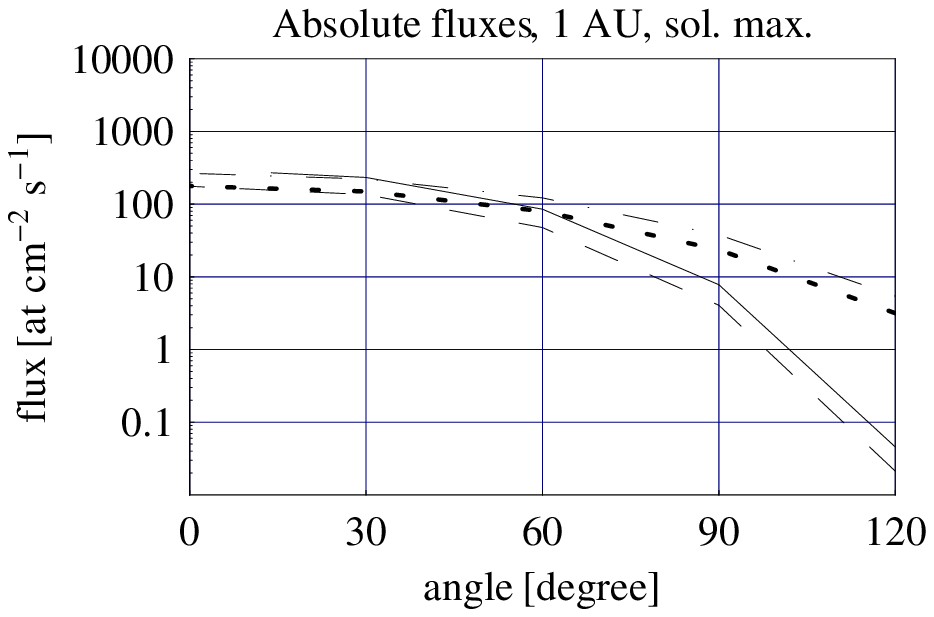}
   \caption{xj Absolute fluxes of neutral interstellar H atoms at 1~AU from the primary and secondary populations for solar minimum (left panel) and solar maximum conditions (right panel), shown as a function of the offset angle from upwind, calculated with the use of Doppler and classical hot models. The classical model results are drawn with solid line (primary populations) and dash-dotted line (secondary population); the Doppler model results with dashed (primary) and dotted lines (secondary). }
              \label{xj}
    \end{figure*}

\begin{figure}
   \centering
   \includegraphics[width=8cm]{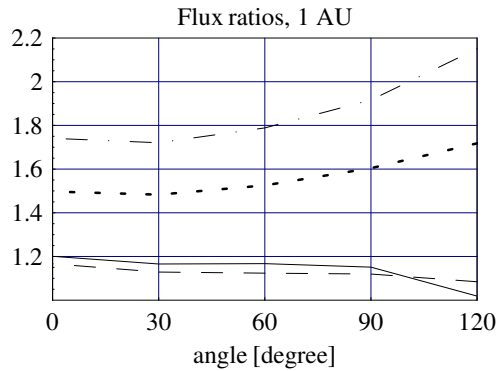}
   \caption{xk Classical/Doppler ratios of the fluxes of primary and secondary populations of neutral interstellar H atoms for solar minimum and maximum conditions as a function of offset angle from upwind. The lower pair of the lines corresponds to solar minimum, the upper pair to solar maximum. Within the solar minimum (lower) pair, the solid line corresponds to the primary population, the broken line to secondary population. Within the solar maximum (upper) pair, the dash-dot line is for primary, the dotted line for secondary population. }
              \label{xk}
    \end{figure}

Even though the differences in the local velocities are small, the effect of the wavelength dependence of radiation pressure on the magnitude of fluxes of the two heliospheric neutral H populations is not negligible because of the appreciable differences in the local densities of the gas. As shown in Fig. \ref{xk}, during solar minimum the classical excess of the flux is 15 to 20\%, depending on the population, and during solar maximum it may reach a factor of 1.4 to 1.9 in the upwind hemisphere and even larger in the downwind region, but with the absolute magnitude of the flux reduced by 2 orders of magnitude from the solar minimum level (Fig. \ref{xj}). In the upwind hemisphere it is very weakly sensitive to the offset angle. The wavelength dependence of radiation pressure must then be appropriately taken into account in interpretation of direct in situ observations of neutral interstellar hydrogen atoms. The local flux of the atoms is a product of the local velocity vector and of the local density, which is directly proportional to the density at the termination shock. Hence the density at the termination shock derived from an in situ-measured flux at 1~AU neglecting  the effect of $v_r$ dependence of radiation pressure will be at least 15\% off the mark (Fig.\ref{xk}), and most probably more, up to a factor of 2.

\begin{figure*}
   \centering
   \includegraphics[width=8cm]{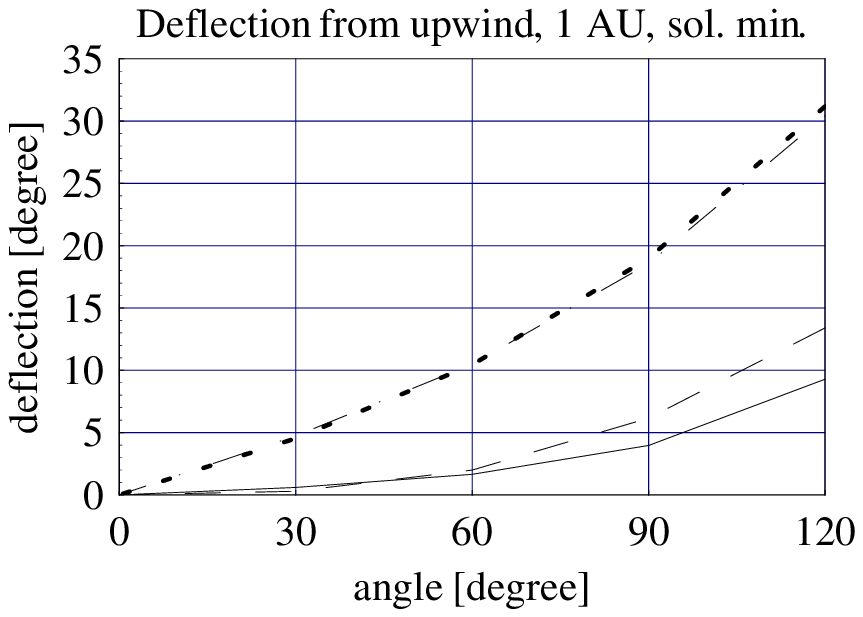}
   \includegraphics[width=8cm]{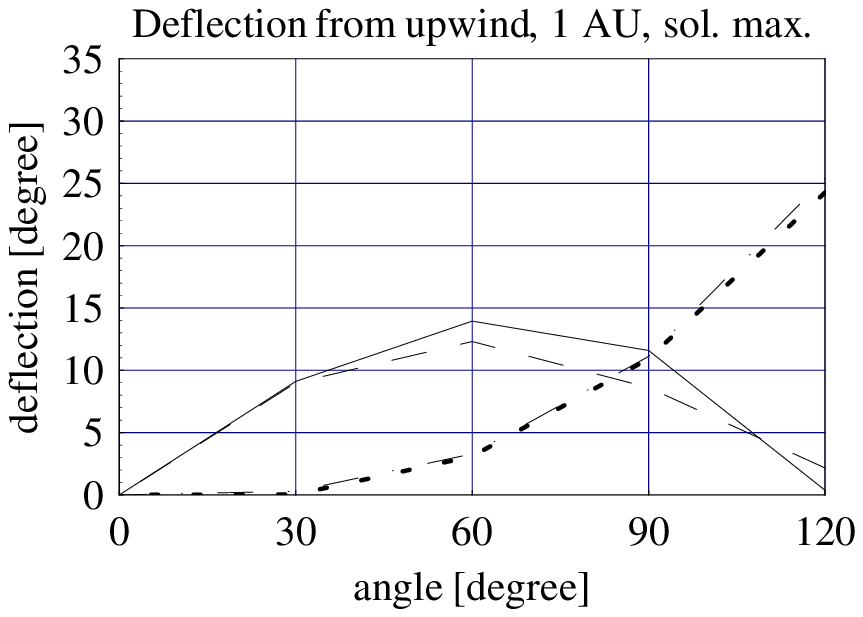}
   \caption{xm Deflection of the local upstream direction of the primary and secondary population of interstellar hydrogen at 1~AU as a function of offset angle from the gas inflow direction for solar minimum and maximum conditions. At solar minimum (left panel), less deflected is the primary population (solid line: classical model, broken line: Doppler model); the secondary is drawn with dotted line (Doppler) and dash-dotted line (classical). Same scheme for solar maximum plot (right panel). }
              \label{xm}
    \end{figure*}

Another factor that potentially can affect interpretation of such measurements are differences in the local bulk velocity vectors of the two populations. \citet{mobius_etal:01a} proposed to use the relative positions of the beams from the primary and secondary populations to check whether their flow directions at the termination shock are parallel or not. Such measurements will hopefully be soon realized by a forthcoming NASA SMEX mission IBEX \citep{mccomas_etal:04a, mccomas_etal:05a, mccomas_etal:06}. Fig. \ref{xm} shows that the deflections of the two populations differ from the predictions of the wavelength-independent model. For both populations, in the region of the Earth orbit where detection by IBEX is the most probable the differences in deflection angle will be on the order of a few degrees, similar, but not identical for both populations. The magnitude of the deflection is a strong function of solar activity, with radiation pressure clearly playing the dominant role; the deflection itself is larger in the case of the Doppler model. 

Even though the differences are just a few degrees, they cannot be regarded as negligible. \citet{quemerais_etal:99} and \citet{lallement_etal:05} suggested an upwind direction of neutral interstellar hydrogen that differs by about $5\degr$ from the inflow direction of interstellar helium measured by \citet{witte:04}. The difference is commonly attributed to a distortion of the heliospheric interface by the extraheliospheric magnetic field. Hence mostly affected should be the secondary population and the difference in the upwind direction of the primary and secondary populations inferred from analysis of in situ measurements interpreted with the use of a $v_r$-insensitive model might be erroneously attributed to a deformation of the interface. Our results show that the deflection of the secondary population obtained from the Doppler and classical models are very similar, but the deflections of the primary population differ by a few degrees.

Finally, we assess the influence of the $v_r$-sensitive radiation pressure on the expected PUI fluxes at 5~AU crosswind (i.e., the location where the H$^+$ PUI flux observed by Ulysses is the strongest).  \citet{bzowski_etal:08a} pointed out that the local production rate of pickup ions, which is a product of the local density of neutral hydrogen and of the local ionization rate, is a weak function of radiation pressure and hence, necessarily, of its details including the $v_r$ dependence. They provided an estimate of the importance of the $v_r$-dependence of radiation pressure for the local production rate of PUI which suggests it is on the level of a few percent. Here we report its influence of on the total flux of PUI, which was computed following the simple approach by \citet{vasyliunas_siscoe:76} and \citet{gloeckler_etal:93a}. Results of these simulations suggest that the classical hot model yields an excess flux with respect to the Doppler model, which at $\sim5$~AU crosswind is equal to only $\sim7.5$\% at solar minimum and to 8.5\% at solar maximum. Thus we conclude that neglection of the $v_r$ dependence of radiation pressure results in an overestimate of the H$^+$ PUI flux at 5~AU from the Sun by about 10\%, irrespective of the solar activity phase.

\subsection{Comparison with the hot model: a scan of the parameter space}
\label{subsec:paramscan}
\begin{figure*}
   \centering
   \includegraphics[width=8cm]{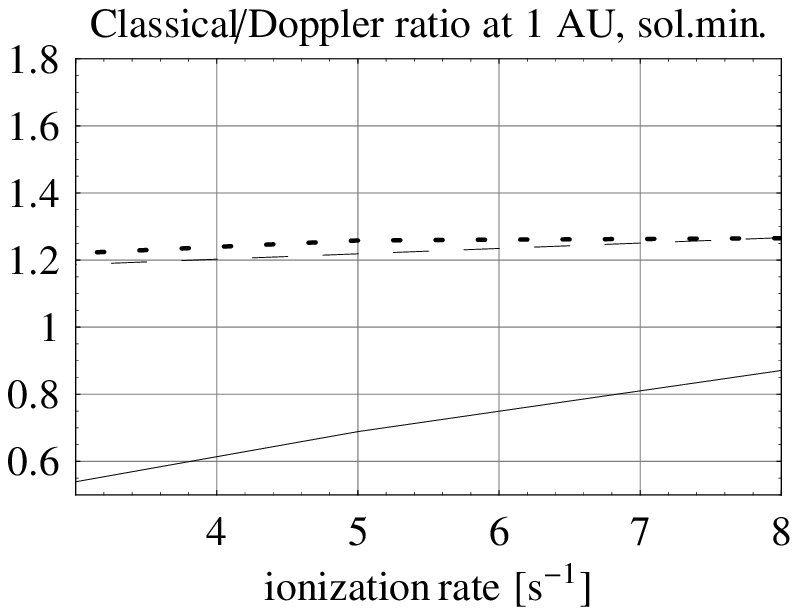}
   \includegraphics[width=8cm]{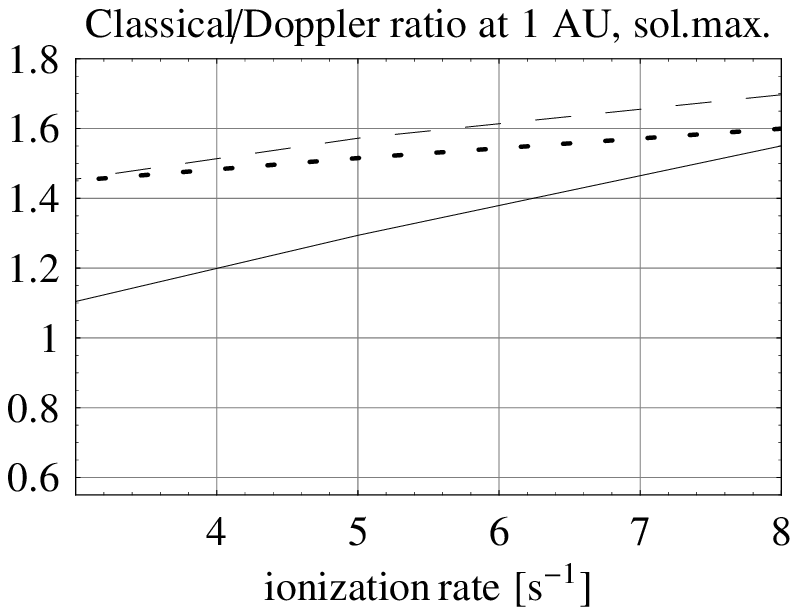}
   \caption{xf Classical hot/Doppler model density ratios for the upwind (dots), crosswind (dashed) and downwind axes (solid) at 1~AU as a function of the ionization rate for solar minimum (left panel) and maximum conditions (right panel). }
              \label{xf}
    \end{figure*}

The density excess $q$ is a weakly increasing and almost linear function of the ionization rate (Fig. \ref{xf}), i.e. the higher the ionization rate one takes, the lower quality of the approximation given by the classical hot model one obtains. This can be explained by preferential elimination of slower atoms from the local ensemble. As discussed by \citet{lallement_etal:85b, bzowski_etal:97}, a result of ionization is a net acceleration of the gas by a few km/s because the atoms that have higher specific velocities in the ensemble preferentially survive against ionization. Such atoms have also higher radial velocities and hence experience a higher radiation pressure due to the non-flat shape of the solar Lyman-$\alpha$ line profile. This force repels them from the Sun stronger than in the case when the solar line profile is flat. As a result, we have a reduction of density -- consequently, the classical model predicts an excess of density, whose magnitude increases with the increase of the ionization rate. 

The sensitivity of the excess to the ionization rate increases with the offset angle from upwind and with the magnitude of the solar flux $I_{\mathrm{tot}}$. At the upwind axis for $I_{\mathrm{tot}}$ corresponding to solar minimum conditions the slope $\partial q/\partial \beta$ is equal to a half of the value relevant for crosswind and to $\sim 0.1$ of the value for downwind. For $I_{\mathrm{tot}}$ relevant for solar maximum conditions the slopes are larger and the highest increase is for the upwind axis. 

In the downwind region, increasing the ionization rate improves the quality of the approximation provided by the classical hot model with respect to the Doppler model. The parameter $q$ is still an increasing function of the ionization rate, but since in the downwind region it is lower than 1, its increase means an improvement. 

\begin{figure*}
   \centering
   \includegraphics[width=8cm]{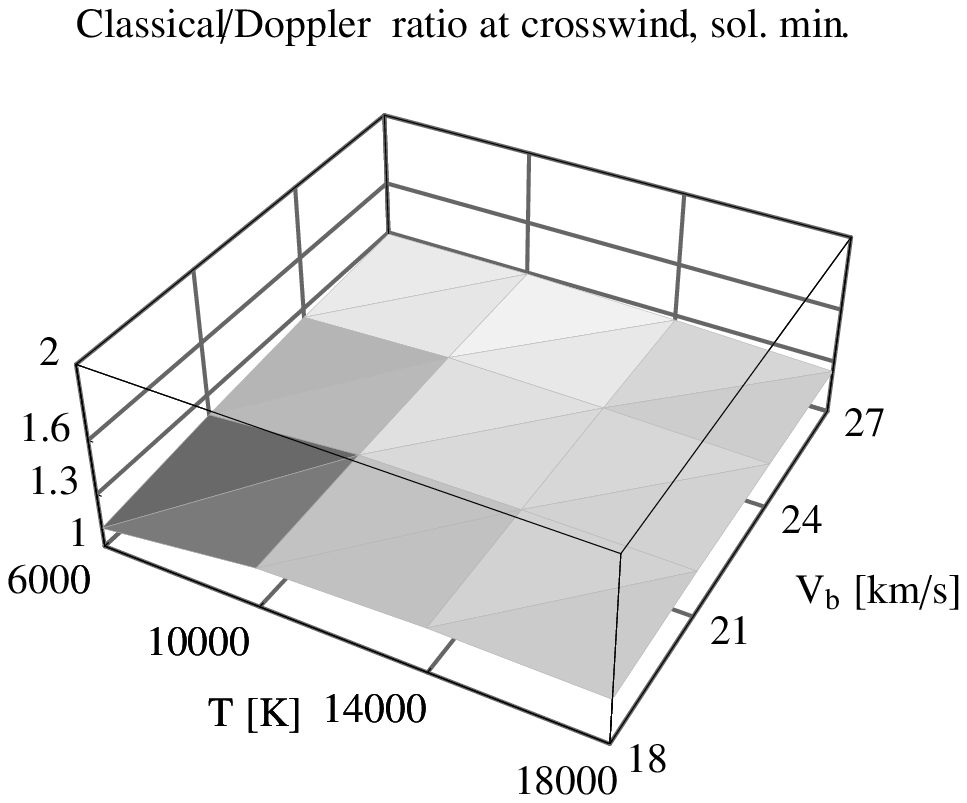}
   \includegraphics[width=8cm]{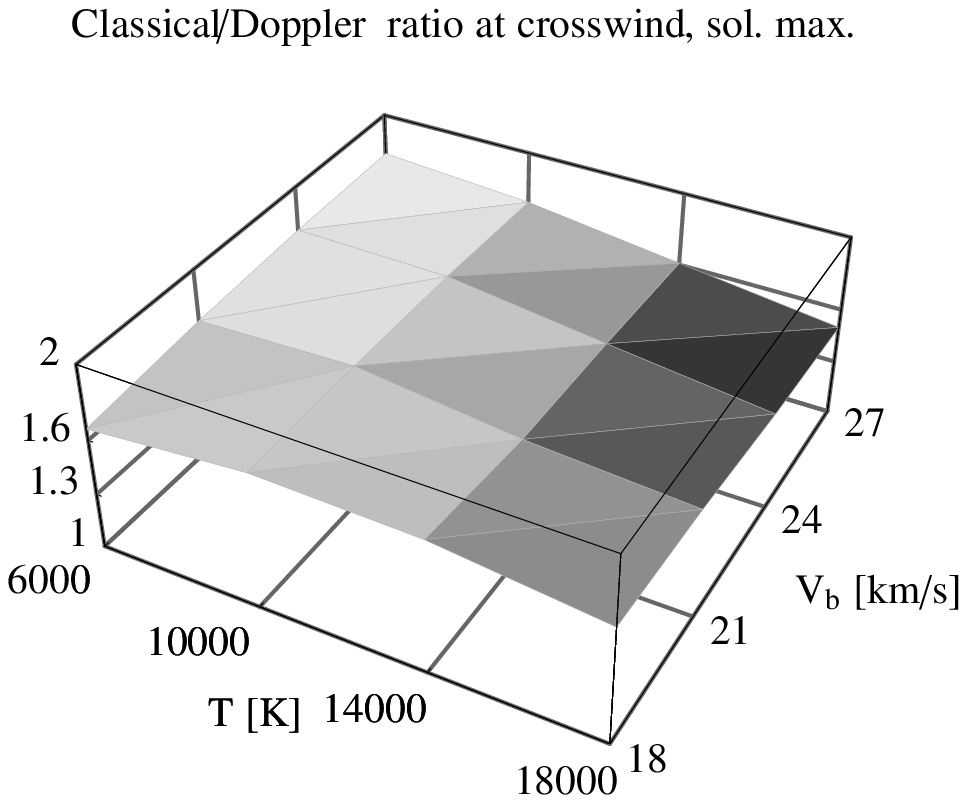}
   \caption{xe Density excess of the classical hot model with respect to the Doppler model of neutral interstellar hydrogen as a function of temperature and bulk speed in the source region for solar minimum (left panel) and solar maximum conditions (right panel). }
              \label{xe}
    \end{figure*}

The magnitude of the excess $q$ is a weak function of temperature and bulk velocity of the atoms at the termination shock, as shown in Fig. \ref{xe}. For $I_{\mathrm{tot}}$ relevant for solar minimum conditions at 1~AU crosswind, it varies from 1.15 to 1.30  depending on the bulk velocity and temperature at the termination shock; for $I_{\mathrm{tot}}$ relevant for solar maximum conditions the amplitude is similar, but the value of $q$ larger.

Generally, the quality of the hot model approximation of density is consistently lower for higher values of $I_{\mathrm{tot}}$, which is modulated by solar activity level. The sensitivity of radiation pressure to radial velocity thus modifies the relations between densities during various phases of solar activity (the amplitude of modulation), discussed, e.g., by \citet{rucinski_bzowski:95a, bzowski_rucinski:95a} and \citet{bzowski_etal:02}.

On the other hand, it seems that the effect on the model values of the heliospheric Lyman-$\alpha$ glow should be on the order of 10\%. An exact calculation requires developing a model, where the local bulk velocity and temperature of the gas and their radial gradients will be taken into account, together with the wavelength dependence of the solar illuminating flux and its wavelength-differential attenuation, which increases with the increase of the solar distance. We believe this effect to be relatively small because during solar minimum, when the contribution to the net backscatter intensity from the gas near the Sun is largest, the $q$ factor is lowest. During solar maximum, when $q$ is largest and hence the quality of the $v_r$-independent model is lowest, the local density of the gas is very much reduced, so increased is the contribution to the net signal of more distant regions of the heliosphere, where the $q$ values are relatively small. In consequence, the $v_r$ dependence of radiation pressure should appreciably affect in situ measurements within $\sim 5$~AU from the Sun, but only in a small degree the photometric observations of the heliospheric glow. 
 
\section{Conclusions}
\label{sec:conclu}

We performed a comparison of density, bulk velocity and flux of neutral interstellar hydrogen in the inner heliosphere, calculated using either the classical hot model or a newly-developed model with radiation pressure being a function of specific radial velocities of individual atoms. The conclusions are the following. 
\begin{enumerate}
\item Differences between the Doppler and classical hot models are restricted to $\sim 5$~AU from the Sun. The classical hot model overestimates the density of neutral hydrogen gas everywhere in the inner heliosphere except a $\sim 60 \degr$ cone around the downwind direction, where a density deficit is predicted. Generally, the density excess/deficit is a strong function of the offset angle from the upwind direction and of the heliocentric distance (Figs \ref{xg}, \ref{xc} and \ref{xd}). 

\item The magnitude of density excess varies with the solar activity level and is higher at solar maximum (Figs \ref{xg} and \ref{xc}), i.e. $v_r$-independent models overestimate the local density stronger during solar maximum.

\item The density excess is a weakly increasing, almost linear function of the ionization rate in the inner heliosphere and a weak almost bi-linear function of the bulk velocity and temperature of the gas at the termination shock (Figs \ref{xf} and \ref{xe}). 

\item The classical model shows a higher upwind-downwind amplitude of radial velocities at 1~AU than the Doppler model; the differences are on the order of 10\%. The excess of absolute flux of neutral atoms returned by the classical model are on the order of 15\% at 1~AU during solar minimum, i.e. at the time of observations by the forthcoming IBEX mission. The excess is much higher during solar minimum, but the absolute values of the fluxes are lower by a few orders of magnitude than during solar minimum (Figs \ref{xi},\ref{xj},\ref{xk},\ref{xl}).

\item The classical model returns a different deflection of the local upstream directions of the primary population of interstellar neutral H than the Doppler model. The deflections of the secondary population returned by the classical and Doppler model are practically identical. Hence interpretation of in situ measurements of these fluxes must be performed very carefully, with appropriate account for the solar line profile shape to avoid confusing deflections caused by the Doppler effect in radiation pressure with, e.g., results of deformation of the heliospheric interface due to external magnetic field (Fig. \ref{xm}). 

\item Discrepancies between the PUI flux at Ulysses in aphelion returned by the Doppler and classical models are on the order of 10\% and are almost insensitive to the phase of solar cycle.
\end{enumerate}
\begin{acknowledgements}
      M.B. gratefully acknowledges the hospitality of the ISSI Institute in Bern, where a part of this work was carried out within the framework of a Working Group on Heliospheric Breathing.
 This work was supported by Polish grants 1P03D00927 and N522~002~31/0902.
\end{acknowledgements}
\bibliographystyle{aa}
\bibliography{iplbib}
\end{document}